\documentclass[a4paper, 10pt]{article}
\usepackage{amsmath, graphicx, xcolor, cite,geometry, multirow, setspace}
\usepackage[utf8]{inputenc}

\newcommand{\rev}{\text{REV}}
\newcommand{\eg}{\textit{e.g.} }

\title{Pressures inside a nano-porous medium. \\The case of a single phase fluid}

\author{Olav Galteland, Dick Bedeaux, and Signe Kjelstrup\\
\small{PoreLab, Department of Chemistry,}\\
\small{Norwegian University of Science and Technology, NTNU}}

\begin{document}
\maketitle

\begin{abstract}
We define the pressure of a porous medium in terms of the grand potential, and compute its value in a nano-confined or nano-porous medium, meaning a medium where thermodynamic equations need be adjusted for smallness. On the nano-scale, the pressure depends in a crucial way on the size and shape of the pores. According to Hill \cite{Hill1964}, two pressures are needed to characterize this situation; the integral pressure and the differential pressure. Using Hill's formalism for a nano-porous medium, we derive an expression for the difference between the integral and the differential pressures in a spherical phase $\alpha$ of radius $R$, $\hat{p}^\alpha-p^\alpha = {\gamma}/{R}$. We recover the law of Young-Laplace for the differential pressure difference across the same curved surface. We discuss the definition of a representative volume element for the nano-porous medium and show that the smallest REV is half a unit cell in the direction of the pore in the fcc lattice. We also show, for the first time,  how the pressure profile through a nano-porous medium can be defined and computed away from equilibrium. 
\end{abstract}
 
\section{Introduction}

The description of transport processes in porous media poses many challenges, well described in the literature, see \eg \cite{Gray2010,bennethum2004,magda1985,todd1995,Ikeshoji2003}. There is, for instance, no consensus, neither on the definition, nor on the measurement or the calculation of the pressure in a porous medium with flow of immiscible fluids. The problem with the ill defined microscopic pressure tensor \cite{todd1995, Hafskjold2002} is accentuated in a heterogeneous system with interfaces between solids and fluids. In a homogeneous fluid phase one may define and calculate a pressure and a pressure gradient from the equation of state. In a porous medium the presence of curved surfaces and fluid confinements makes it difficult to apply accepted methods for calculation of the microscopic pressure tensor and, consequently, the pressure gradient as driving force for fluid flow. The scale at which we choose to work will be decisive for the answer. Moreover, the scale that the hydrodynamic equations of transport refer to, remains to be given for nano-porous as well as micro-porous media.  

A central element in the derivation of the equations of transport on the macro-scale is the definition of a representative volume element (REV), see e.g. \cite{Hassanizadeh1990,Gray1998}. The size of the REV should be large compared to the pore-size and small compared to size of the porous medium. It should contain a statistically representative collection of pores. We have recently discussed \cite{Kjelstrup2018a} a new scheme to define as basis set of additive variables: the internal energy, entropy, and masses of all the components of the REV. These variables are additive in the sense that they are sums of contributions of all phases, interfaces and contact lines within the REV. Using Euler homogeneity of the first kind, we were able to derive the Gibbs equation for the REV. This equation defines the temperature, pressure and chemical potentials of the REV as partial derivatives of the internal energy of the REV \cite{Kjelstrup2018a}. 

As discussed in \cite{Kjelstrup2018b} the grand potential, $\Upsilon$, of the REV is given by minus $k_\text{B}T$ times the logarithm of the grand partition function, $Z_\text{g}$, where $k_\text{B}$ is Boltzmann's constant and $T$ is the temperature. The grand potential is equal to minus the contribution to the internal energy from the pressure-volume term, $k_\text{B}T\ln Z_\text{g}= \Upsilon = -pV$, which we will from now on refer to as the compressional energy. For a single fluid $f$ in a porous medium $r$, the result was \cite{Kjelstrup2018a,Kjelstrup2018b}
\begin{equation}
  pV=p^{f}V^{f}+p^{r}V^{r}-\gamma ^{fr}\Omega ^{fr},  
  \label{eq:pV}
\end{equation}
where $p$ and $V$ are the pressure and the volume of the REV. Furthermore $p^{f}$ and $V^{f}$ are the the pressure and the volume of the fluid in the REV, $p^{r}$ and $V^{r}$ are the the pressure and the volume in the grains in the REV, and $\gamma ^{fr}$ and $\Omega ^{fr}$ are the surface tension and the surface area between the fluid and the grain. 
The assumption behind the expression was the additive nature of the grand potential. This definition of the REV, and the expression for the grand potential opens up a possibility to define the pressure on the hydrodynamic scale. The aim of this work is to explore this possibility.  We shall find that it will work very well for flow of a single fluid in a porous medium. As a non-limiting illustrative example, we use grains positioned in a fcc lattice. 
The work can be seen as a continuation of our earlier works \cite{Kjelstrup2018a,Kjelstrup2018b}.  

The work so far considered transport processes in micro-porous, not nano-porous media. In micro-porous media, the pressure of any phase (the surface tension of any interface) is independent of the volume of the phase (the area between the phases). This was crucial for validity of equation \ref{eq:pV}. For nano-porous systems, we need to step away from equation \ref{eq:pV}. Following Hill's procedure for small systems' thermodynamics \cite{Hill1964}, we generalize equation \ref{eq:pV} to provide an expression for the thermodynamic pressure in a nano-porous medium. We shall see that not only one, but two pressures are needed to handle the additional complications that arise  at the nano-scale; the impact of confinement and of radii of curvature of the interfaces. 
In the thermodynamic limit, the approach presented for the nano-scale must simplify to the one for the macro-scale. We shall see that this is so. In order to work with controlled conditions, we will first investigate the pressure of a fluid around a single solid nano-scale grain and next around a lattice of solid nano-scale grains. The new expression, which we propose as a definition of the pressure in a nano-porous medium, will be investigated for viability and validity for this case. The present work can be seen as a first step in the direction towards a definition and use of pressure and pressure gradients in real porous media. 

The pressure is not uniquely defined at molecular scale. This lack of uniqueness becomes apparent in molecular dynamics (MD) simulations, for which the compuational algorithm has to be carefully designed \cite{Hafskjold2002}. The predominant method for pressure calculations in particular systems is using the Irving-Kirkwood contour for the force between two particles \cite{Irving1950}. This algorithm works for homogeneous systems, but special care must be taken for heterogeneous systems \cite{todd1995,Ikeshoji2003}. However, if the control volume (REV) used for pressure calculation is large compared with the heterogeneity length scale, one may argue that  the the algorithm for homogeneous systems gives a good approximation to the true result. We are interested in the isotropic pressure averaged over the REV, on a scale where the porous medium can be considered to be homogeneous.

The paper is organized as follows. In section 2 we derive the pressure of a REV for one solid grain surrounded by fluid particles (Case I) and for a three-dimensional face-centered cubic (fcc) lattice of solid grains (Case II). Section 3 describes the molecular dynamics simulation technique when the system is in equilibrium and in a pressure gradient. In section 4 we use the theory to interpret results of equilibrium molecular dynamics simulations for one solid grain and for an array of solid grains in a fluid. Finally we apply the results to describe the system under a pressure gradient. We conclude in the last section that the expressions and the procedure developed provide a viable definition of the pressures and pressure gradients in nano-porous media.

\section{The pressure of a nano-porous medium}

Equation \ref{eq:pV} applies to a micro-porous medium, a medium where the pore-size is in the micrometer range or larger \cite{Kjelstrup2018a,Kjelstrup2018b}. For a nano-porous medium we need to apply the thermodynamics of small systems \cite{Hill1964}. In nano-porous media, this technique is therefore well suited for the investigation. The thermodynamic properties like internal energy, entropy and masses of components of a small system are not proportional to the system's volume. As Hill explained, this leads to the definition of two different pressures, for which he introduced the names integral and differential pressure, $\hat{p}$ and $p$, respectively. For a system with a volume $V$, these pressures are related by
\begin{equation}
  p(V) = \frac{\partial \left( \hat{p}\left( V\right) V\right) }{\partial V}
  = \hat{p}\left( V\right) +V\frac{\partial \left( \hat{p}\left( V\right)\right) }{\partial V}.  
  \label{eq:p_hill}
\end{equation}
The symbol $p$ (the differential pressure) is given to the variable that we normally understand as the pressure on the macroscopic level. It is only when $\hat{p}$ depends on $V$, that the two pressures are different. For large systems,  $\hat{p}$ does not depend on $V$ and the two pressures are the same. 

The integral and differential pressures connect to different types of mechanical work on the ensemble of small systems.  The differential pressure times the change of the small system volume is the work done on the surroundings by this volume change. The name differential derives from the use of a differential volume. This work is the same, whether the system is large or small. The integral pressure times the volume per replica is, however, the work done by adding one small system of constant volume to the remaining ones, keeping the temperature constant.  This work is special for small systems. It derives from the ensemble view, but is equally well measurable.   The word integral derives from the addition of a small system.

From statistical mechanics for macro-scale systems, we know that $pV=\hat{p}V$ equals $k_{\text{B}}T$\ times the natural logarithm of the grand-canonical partition function. For a small (nano-sized) system, Hill (\cite{Hill1964}, equation 1-17), showed that this logarithm gives $\hat{p}V$. In nano-porous media this product is different from $pV$, cf. equation \ref{eq:p_hill}. Energies are still additive and the total compressional energy within the small system is similar to equation \ref{eq:pV}.  We replace equation \ref{eq:pV} by: 
\begin{equation}
  \hat{p}V=\hat{p}^{f}V^{f}+\hat{p}^{r}V^{r}-\hat{\gamma}^{fr}\Omega ^{fr},
  \label{eq:p_hatV}
\end{equation}
where $\hat{p}^{f},\ \hat{p}^{r}$ are integral pressures of the sub-volumes $V^{f}$\ and $V^{r}$, and $\hat{\gamma}^{fr}$ is the integral surface tension.

We consider here a nano-porous medium, so integral pressures and integral surface tensions apply. The integral pressure and integral surface tension normally depend on the system size. In the porous medium there are two characteristic sizes: the size of a grain and the distance between the surfaces of two grains.\footnote{Another valid characteristic size is the size of the pores between the grains, but this follows from the two we have chosen.} The quantities $\hat{p}$, $\hat{p}^{f}$, $\hat{p}^{r} $ and $\hat{\gamma}^{fr}$ may depend on both. We shall here examine a system (cf. section 3) of spherical, monodisperse grains, for which the radius $R$ is a good measure of the size. The volume of the grains may be a good alternative measure, which we will also use. The dependence on the grain size and on the distance between the surfaces of the grains will be studied in an effort to establish equations \ref{eq:pV} and \ref{eq:p_hatV}.

In the following, we consider a single spherical grain confined by a single phase fluid (Case I) and a face-centered cubic (fcc) lattice of spherical grains confined by a single phase fluid (Case II). The size of the REV does not need to be large, and we will show in section 4.2 that the smallest REV is half a unit cell in the direction of the pore in the fcc lattice.

\subsection{Case I. Single spherical grain}
\label{case1}

\begin{figure}
  \centering
  \includegraphics[width=0.6\textwidth]{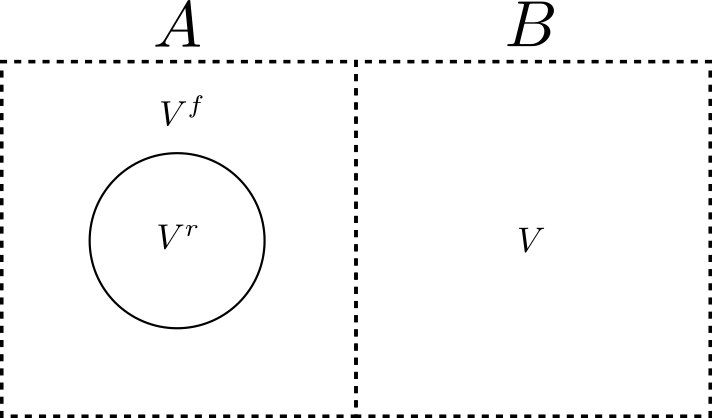}
  \caption{A particle in a confined system (A) in equilibrium with a bulk fluid phase (B).}
  \label{fig:system}
\end{figure}

Consider the spherical inclusion of a grain $r$ in a box with a fluid phase $f$. This is system A of figure \ref{fig:system}. Phase $f$ has the volume $V^{f}$ and phase $r$ has the volume $V^{r}$. The total volume is $V=V^{f}+V^{r}$. The surface area between phases $f$ and $r$ is $\Omega ^{fr}$. System A is in equilibrium with system B, which has the same volume $V$, and contains fluid $f$ only.

We now make the assumptions that $\hat{p}^{r}$ depends only on $V^{r}$, while $\hat{p}$, $\hat{p}^{f}$ and $\hat{\gamma}^{fr}$ do not depend on $V^{r}$ or any other size variable. This means that the spherical grain $r$ is considered to be small in Hill's sense, while the fluid phase is a large thermodynamic system. When the surface tension depends on the curvature, there is a dependence of $\hat{\gamma}^{fr}$ on $\Omega^{fr}$ \cite{Tolman1949,Helfrich1973}. This interesting effect, which we will not consider here, becomes relevant as the grain size decreases. The assumptions will be used to find the contributions to equation \ref{eq:p_hatV}. The assumptions mean that
\begin{equation}
  \hat{p}^r\neq p^r, \quad \hat{p}=p, \quad 
  \hat{p}^{f}=p^{f} \quad \text{and} 
  \quad \hat{\gamma}^{fr}=\gamma^{fr}.
  \label{eq:assumptions}
\end{equation}
Equation \ref{eq:p_hatV} for system A in figure \ref{fig:system} reduces to 
\begin{equation}
  pV=p^{f}V^{f}+\hat{p}^{r}V^{r}-\gamma ^{fr}\Omega ^{fr}.  \label{eq:p_hatV2}
\end{equation}

An interesting implication of $\hat{p}^f=p^f$ is that we can use the standard MD algorithms as implemented for large systems to calculate $p^f$ in the fluid phase. It also implies that in equilibrium, $p^{fA}=p^{fB}$ as $p^f$ has to be continuous on the border between A and  B.

To proceed, we examine the state where system A in figure \ref{fig:system} is in equilibrium with system B. The phase $f$ of system B has the pressure $p^{f}$. The equilibrium condition that makes the pressure of system A equal to the pressure of system B, is
\begin{equation}
  p=p^{f}.
  \label{eq:equil}
\end{equation}
By introducing this into equation \ref{eq:p_hatV2}, we obtain the integral pressure inside $V^{r}$ at equilibrium,
\begin{equation}
  \hat{p}^{r}=p^{f}+\gamma ^{fr}\frac{\Omega ^{fr}}{V^{r}}=p^{f}+\frac{3\gamma
  ^{fr}}{R}.
  \label{1.3}
\end{equation}
where we used that $V^{r}=4\pi R^{3}/3$ and $\Omega ^{fr}=4\pi R^{2}$ for a sphere of radius $R$. From the definition of Hill \cite{Hill1964} the differential pressure is, cf. equation \ref{eq:p_hill},
\begin{eqnarray}
  p^{r} &=&\frac{\partial (\hat{p}^{r}V^{r})}{\partial V^{r}}=\frac{\partial
  (p^{f}V^{r})}{\partial V^{r}}+\gamma ^{fr}\frac{\partial \Omega ^{fr}}{%
  \partial V^{r}}  \notag \\
  &=&p^{f}+\gamma ^{fr}\frac{\partial \Omega ^{fr}}{\partial V^{r}}=p^{f}+%
  \frac{2\gamma ^{fr}}{R},  \label{1.4}
\end{eqnarray}
where we used that
\begin{equation}
  \frac{\partial \Omega ^{fr}}{\partial V^{r}}=\frac{\partial \Omega ^{fr}}{\partial R}\left( \frac{\partial V^{r}}{\partial R}\right) ^{-1}
  =
  \frac{2}{R}.
  \label{eq:dOdV}
\end{equation}
We see that the differential pressure inside the grain satisfies Young-Laplace's law for pressure differences across curved interfaces. The fact that we recover this well-known law, supports the validity of the assumptions made. 

By subtracting equation \ref{1.4} from equation \ref{1.3}, we obtain an interesting new relation
\begin{equation}
  \hat{p}^r-p^r = \frac{\gamma^{fr}}{R}
  \label{eq:newrelation}
\end{equation}
The expression relates the integral and differential pressure for a spherical phase $r$ of radius $R$. It is clear that this pressure difference is almost equally sensitive to the radius of curvature as is the pressure difference in Young-Laplace's law.

We also see from this example how the integral pressure enters in the description of small systems. The integral pressure is not equal to our normal bulk pressure, called the differential pressure by Hill, $\hat{p}^{r}\neq p^{r}$. While two differential pressures satisfy Young-Laplace's law, the integral pressure does not! The integral pressure has the property that when averaged over system A using equation \ref{eq:p_hatV2}, it is the same as in system B, cf. equation \ref{eq:equil}. This analysis shows that system A is a possible, or as we shall see proper, choice of a REV that contains the solid grain, while system B is a possible choice of a REV that contains only fluid. 

An alternative way to derive equation \ref{eq:equil} is given in the Appendix.

\subsection{Case II. Lattice of spherical grains}

The above explanation concerned a single spherical grain, and was a first step in the development of a procedure to determine the pressure of a nano-porous medium. To create a more realistic model, we introduce now a lattice of spherical grains. The integral pressure of a REV containing many grains is given by an extension of equation \ref{eq:p_hatV} 
\begin{equation}
  \hat{p}V=\hat{p}^{f}V^{f}+\sum_{i}^n\hat{p}_{i}^{r}V_{i}^{r}-\sum_{i}^n\hat{%
  \gamma}_{i}^{fr}\Omega_{i}^{fr},  
  \label{eq:p_hatV_fcc}
\end{equation}
where $n$ is the number of grains in the volume. For each grain one may follow the same derivation for the integral and differential pressure as for the single grain. The equilibrium condition is
\begin{equation}
  \hat{p}=\hat{p}^{f}=p^{f}.
\end{equation}
By using equation \ref{1.3}, we obtain
\begin{equation}
  \hat{p}_{i}^{r}=p^{f}+\gamma _{i}^{fr}\frac{\Omega _{i}^{fr}}{V_{i}^{r}}%
  =p^{f}+\frac{3\gamma _{i}^{fr}}{R_{i}},  
  \label{1.7}
\end{equation}
where the last identity applies to spherical grains only. The differential pressure of the grains is given by a generalization of equation \ref{1.4}%
\begin{eqnarray}
  p_{i}^{r} &=&\frac{\partial (\hat{p}_{i}^{r}V_{i}^{r})}{\partial V_{i}^{r}}=%
  \frac{\partial (p^{f}V_{i}^{r})}{\partial V_{i}^{r}}+\gamma _{i}^{fr}\frac{%
  \partial \Omega _{i}^{fr}}{\partial V_{i}^{r}}  \notag \\
  &=&p^{f}+\gamma _{i}^{fr}\frac{\partial \Omega _{i}^{fr}}{\partial V_{i}^{r}}%
  =p^{f}+\frac{2\gamma _{i}^{fr}}{R_{i}},  \label{1.8}
\end{eqnarray}
where the last identity is only for spherical grains. The differential pressures again satisfy Young-Laplace's law.

When all grains are identical spheres and positioned on a fcc lattice, a properly chosen layer covering half the unit cell can be a proper choice of the REV. We shall see how this can be understood in more detail from the molecular dynamics simulations below.

\section{Molecular dynamics simulations}

Cases I and II were simulated at equilibrium, while Case II was simulated also away from equilibrium. Figures \ref{fig:sphere} and \ref{fig:several_spheres} illustrate the equilibrium simulations of the two cases.

\begin{figure}
  \centering
  \includegraphics[width=1.0\linewidth]{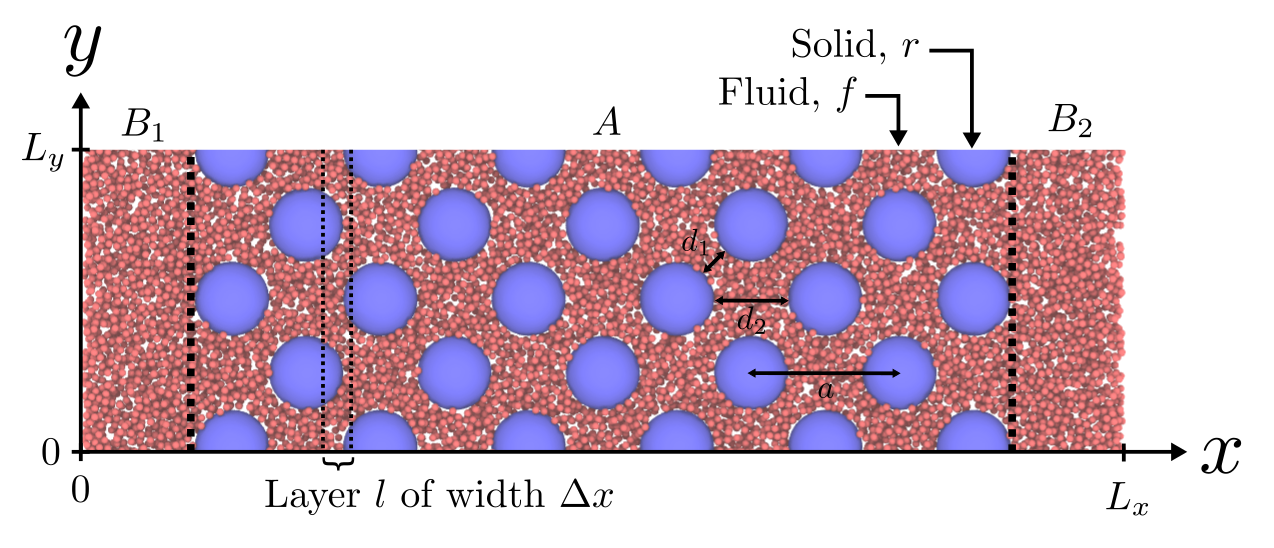}
  \caption{A slice of the simulation box in case II. The box has side lengths $L_x, L_y, L_z$, and properties are calculated along the $x$-axis in layers $l$ of width $\Delta x$. Blue particles are grain $r$ and red particles are fluid $f$. The A is the lattice constant of the fcc lattice, $d_1$ and $d_2$ are the two shortest surface-to-surface distances.}
  \label{fig:explained_system}
\end{figure}

\subsection{Systems}

The simulation box was three-dimensional with side lengths $L_{x},L_{y},L_{z}$, where it was elongated in the $x$-direction, $L_x > L_y=L_z$. Periodic boundary conditions were used in all directions in the equilibrium simulations. In the non-equilibrium simulation reflecting particle boundaries \cite{Li1998} were applied to the $x$-direction, cf. section \ref{sec:neq}. Along the $x$-axis, the simulation box was divided into $n$ rectangular cuboids (called layers) of size $\Delta x,L_{y},L_{z}$, where $\Delta x=L_{x}/n$. The volume of each layer is $V_{l}=\Delta xL_{y}L_{z}$. There are two regions A and B in the simulation box. Region A contains fluid (red particles) and grain (blue particles) and region B contains only fluid, see figure \ref{fig:explained_system}. The regions, $B=B_{1}+B_{2}$ and A do not have the same size, but the layers have the same thickness, $\Delta x$. The compressional energy of the fluid in one layer is, $\hat{p}_{l}^{f}V_{l}^{f}=p_{l}V_{l}^{f}$.

The simulation was carried out with LAMMPS \cite{Plimpton1995} in the canonical ensemble using the Nosé-Hoover thermostat \cite{Hoover1996}, at constant temperature $T^*=2.0$ (in Lennard-Jones units). The critical temperature for the Lennard-Jones/spline potential (LJ/s) is approximately $T^*_c\approx 0.9$. Fluid densities range from $\rho^*=0.01$ to $\rho^*=0.7$.

\subsection{Case studies} 

In Case I the single spherical grain was placed in the center of the box. A periodic image of the spherical grain is a distance $L_x$, $L_y$ and $L_z$ away in the $x$, $y$ and $z$-directions, see figure \ref{fig:sphere}a. The surface to surface distance of the spherical grains is $d=L_\alpha - 2R$, where $R$ is the radius of the grain, and $\alpha = y,z$.  In Case I, each spherical grain has four nearest neighbours in the periodic lattice that is built when we use periodic boundary conditions. We considered two nearest neighbour distances;  $d=4\sigma_0$ and $d=11\sigma_0$, where $\sigma_0$ is the diameter of the fluid particles. 

\begin{figure}
  \centering
  \includegraphics[width=0.7\linewidth]{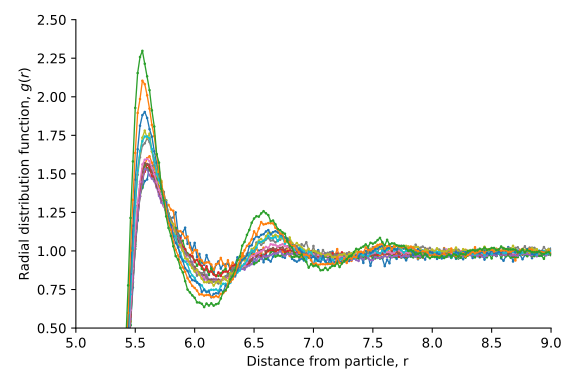}
  \caption{The radial distribution function of fluid particles around a grain, as shown in figure \ref{fig:sphere}. Results are shown for densities that vary between $\rho^*=0.1$ and $\rho^*=0.7$.}
  \label{fig:rdf}
\end{figure}

In case II, the spherical grains were placed in a fcc lattice with lattice constant A. The two shortest distances between the surfaces were characterized by  $d_1 = \frac{1}{2}(\sqrt{2}a-4R)$ and $d_2 = a-2R$, see figure \ref{fig:explained_system}, where $d_1<d_2$. We used $d_1 = 4.14\sigma_0$ and $d_1 =11.21\sigma_0$, which is almost the same as the distances considered in Case I. The corresponding other distances were  $d_2 = 10\sigma_0$ and $d_2 =20\sigma_0$. Each grain has 12 nearest neighbours at a distance $d_1$. The radial rock-fluid pair correlation function was computed for various fluid densities, see figure \ref{fig:rdf}.  The radius of the particle is $R=5.4 \sigma_0$. 

In all cases we computed the volume of the grains $V_{l}^{r}$, the surface area $\Omega _{l}^{fr}$ and the compressional energy of each layer, $l$, in the $x$-direction.  

\subsection{Particle interaction potential}

The particles interact with the Lennard-Jones/spline potential, 
\begin{equation}
u_{ij}(r)=%
  \begin{cases}
  \infty  & \text{if }r<R_{ij} \\ 
  4\epsilon _{ij}\left[ \left( \frac{\sigma _{ij}-R_{ij}}{r-R_{ij}}\right)
  ^{12}-\left( \frac{\sigma _{ij}-R_{ij}}{r-R_{ij}}\right) ^{6}%
  \right]  & \text{if }R_{ij}<r<r_{s,ij} \\ 
  a_{ij}(r-r_{c,ij})^{2}+b_{ij}(r-r_{c,ij})^{3} & \text{if }r_{s}<r<r_{c,ij}
  \\ 
  0 & \text{if }r>r_{c,ij}%
  \end{cases}.  
  \label{2.1}
\end{equation}
Each particle type has a hard-core diameter $R_{ii}$ and a soft-core diameter $\sigma _{ii}$. There were two types of particles, small particles with $\sigma _{ff}=\sigma _{0},\ R_{ff}=0$ and large particles with $\sigma_{rr}=10\sigma _{0},\ R_{rr}=9\sigma _{0}$. The small particles are the fluid ($f$), and the large particles are the grain ($r$). The hard-core and soft-core diameters for fluid-grain pairs are given by the Lorentz mixing rule
\begin{equation}
  R_{fr}=\frac{1}{2}\left( R_{ff}+R_{rr}\right) \quad \text{and} \quad
  \sigma_{fr}=\frac{1}{2}\left( \sigma _{ff}+\sigma _{rr}\right).
  \label{2.2}
\end{equation}
We define the radius of the grain particles as $R\equiv ({\sigma _{ff} +\sigma_{rr}})/{2}= 5.5\sigma_0$, which is the distance from the grain center where the potential energy is zero. This implies that fluid particles can occupy a position closer to the grain than this, especially at high pressures, this is shown in figure \ref{fig:rdf}.

The interaction strength $\epsilon_{ij}$ was set to $\epsilon_0$ for all particle-particle pairs. The potential and its derivative are continuous in $r=r_{c,ij}$. The parameters $a_{ij}$, $b_{ij}$ and $r_{s, ij}$ are determined so that the potential and the derivative of the potential (the force) are continuous at $r=r_{s,ij}$.

\subsection{Pressure computations}

The contribution of the fluid to the grand potential of layer $l$ is \cite{Irving1950}
\begin{equation}
  p_{l}^{f}V_{l}^{f} =
  \frac{1}{3}\left\langle \sum_{i \in l }m_i (\textbf{v}_i\cdot\textbf{v}_i)\right\rangle
  -\frac{1}{6}\left\langle \sum_{i\in l}\sum_{j=1}^N(\mathbf{r}_{ij}\cdot \mathbf{f}_{ij})\right\rangle ,  
  \label{2.3}
\end{equation}
where $p_{l}^{f}$ is the fluid integral pressure, $V_{l}^{f}$ the fluid volume, $m_i$ and $\textbf{v}_i$ are the mass and velocity of fluid particle $i$. The first two sums are over all fluid particles $i$ layer $l$, while the second sum is over all other particles $j$. $\mathbf{r}_{ij}\equiv  \mathbf{r}_{i}-\mathbf{r}_{j}$ is the vector connecting particle $i$ and $j$, and $\mathbf{f}_{ij}=-\partial u_{ij}/\partial \mathbf{r}_{ij}$ is the force between them. The $\cdot$ means a inner product of the vectors. The computation gives $\hat{p}_{l}^{f}$ which is the contribution to the integral pressure in layer $l$ from the fluid particles, accounting for their interaction with the grain particles. 

\subsection{The porous medium in a pressure gradient}
\label{sec:neq}
We used the reflecting particle boundary method developed by Li \textit{et al.} \cite{Li1998} to generate a pressure difference across the system along the $x$-axis. Where particles pass the periodic boundary at $x=0$ and $x=L_x$ with probability $\left(1-\alpha_p \right)$ and reflected with probability $\alpha_p$. A large $\alpha_p$ gives a high pressure difference and a low $\alpha_p$ gives a low pressure difference. 

\section{Results and discussion}

The results of the molecular dynamics simulations  are shown in figures \ref{fig:sphere}-\ref{fig:compare} (equilibrium) and figures \ref{fig:pressure_gradient}, \ref{fig:pressure_gradient_smooth} (away from equilibrium). The porous medium structure was characterized by its pair correlation function, cf. figure \ref{fig:rdf}. The compressional energy was computed according to equation \ref{eq:p_hatV2} in case I with a single spherical grain and case II with a lattice of spherical grains. 

We computed the compressional energy, $p_{l}V_{l}$, in the bulk liquid (region B) and in the nano-porous medium (region A). In the bulk liquid we computed the pressure directly from the compressional energy, 
because $p_lV_l = p_l^fV_l^f$ (not shown). 

Figures \ref{fig:sphere} and \ref{fig:several_spheres} show the contributions from the solid phase $\hat{p}_{l}^{r}V_{l}^{r}$ and fluid-solid surface $\gamma _{l}^{fr}\Omega _{l}^{fr}$, cf. equation \ref{eq:p_hatV2}. The grain particles were identical and the system was in equilibrium, so the integral pressure in the grains was everywhere the same, $\hat{p}_{l}^{r}=\hat{p}^{r}$. 
Similarly, the surface tension was everywhere the same, $\gamma_{l}^{fr}=\gamma ^{fr}$. 

In figures \ref{fig:caseI}, \ref{fig:caseII} and \ref{fig:compare} the grain pressure $\hat{p}^{r}$ and surface tension $\gamma^{fr}$ are plotted as a function of the fluid pressure $p^f$. The results for Case II were next used in figures \ref{fig:pressure_gradient} and \ref{fig:pressure_gradient_smooth} to determine the pressure gradient across the sequence of REVs in the porous medium. 

\subsection{Case I. Single spherical grain. Equilibrium}

\begin{figure}
  \centering
  \includegraphics[width=0.7\linewidth]{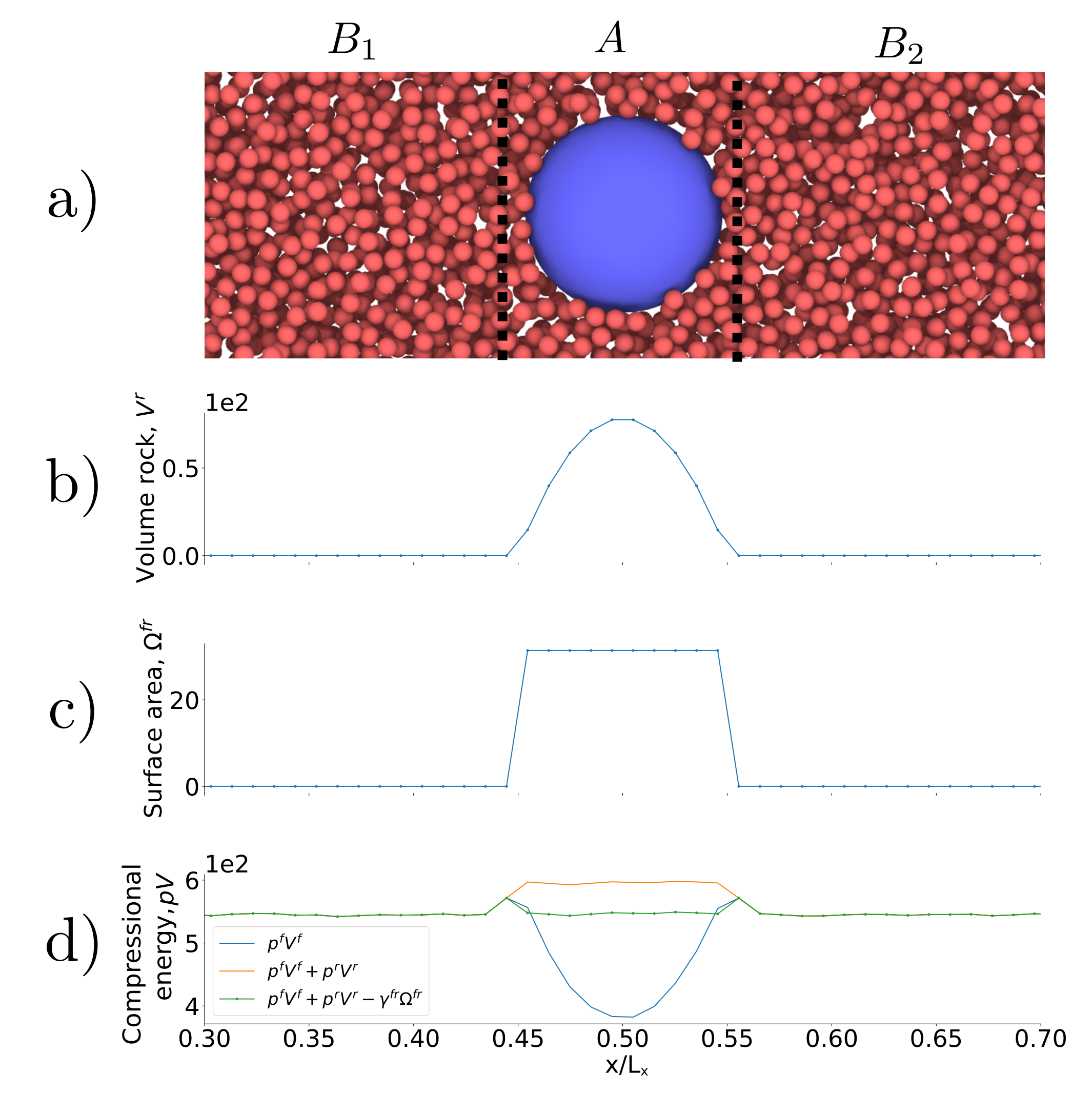}
  \caption{\textbf{a)} Illustration of case I, a single spherical grain surrounded by a fluid phase with $d=4\sigma_0$. \textbf{b)} Volume of grain, $V^r$, \textbf{c)} surface area $\Omega^{fr}$ and \textbf{d)} compressional energy $pV$ as a function of the $x$-axis of the simulation box.}
  \label{fig:sphere}
\end{figure}

The single sphere case is illustrated in figure \ref{fig:sphere}a. 
Figures \ref{fig:sphere}b and \ref{fig:sphere}c show the variation in the volume of the porous medium (rock), $V_l^r$, and the surface area between the rock and the fluid, $\Omega^{fr}$, along the $x$-axis of the simulation box. The two quantities were determined for all layers, $l$, and these results were used in the plots of figure \ref{fig:sphere}b and \ref{fig:sphere}c. To be representative, the REV must include the solid sphere with boundaries left and right of the sphere. In order to obtain $p^\rev V^\rev$ we summed $p_{l}V_{l}$ over all the layers in the REV. 
At equilibrium, $p^\rev=p$, where $p$ is the pressure in the fluid in region B. For the REV we then have
\begin{equation}
  pV^\rev= \sum_{l\in \rev} p_{l}^{f}V_{l}^{f}+\hat{p}^{r}\sum_{l\in \rev}V_l^r-\gamma
  ^{fr}\sum_{l\in \rev}\Omega _{l}^{fr},  \label{R.1}
\end{equation}
where we used that $\hat{p}_{l}^{r}=\hat{p}^{r}$ and $\gamma_{l}^{fr}=\gamma ^{fr}$. We know the values of all the elements in this equation, except $\hat{p}^{r}$ and $\gamma ^{fr}$. The values of $\hat{p}^{r}$ and $\gamma ^{fr}$ are next chosen such that equation \ref{R.1} holds. With these fitted values available, we calculated $p_{l}V_{l}$ of each layer from 
\begin{equation}
  p_{l}V_{l}=p_{l}^{f}V_{l}^{f}+\hat{p}^{r}V_{l}^{r}-\gamma^{fr}\Omega _{l}^{fr}.
  \label{R.2}
\end{equation}
The contributions to the compressional energy in this equation for Case I are shown in the bottom figure \ref{fig:sphere}d. We see the contribution from (1) the bulk fluid $p_{l}^{f}V_{l}^{f}$, (2) the bulk fluid and grain $p_{l}^{f}V_{l}^{f}+\hat{p}^{r}V_{l}^{r}$ and (3) the total compressional energy,  $p_{l}V_{l}=p_{l}^{f}V_{l}^{f}+\hat{p}^{r}V_{l}^{r}-\gamma^{fr}\Omega^{fr}$, which gives the pressure of the REV when summed and divided with the volume of the REV.    

Figure \ref{fig:sphere}d shows clearly, that the bulk pressure energy gives the largest contribution, as one would expect. It is also clear that the surface energy is significant.  As the surface to volume ratio increases, the bulk contributions may become smaller than the surface contribution. In the present case, this will happen when the radius of the sphere is  $2.25 \sigma_0$. For our grains with $R=5.5 \sigma_0$ this does not happen. 

\begin{figure}
  \centering
  \includegraphics[width=0.85\linewidth]{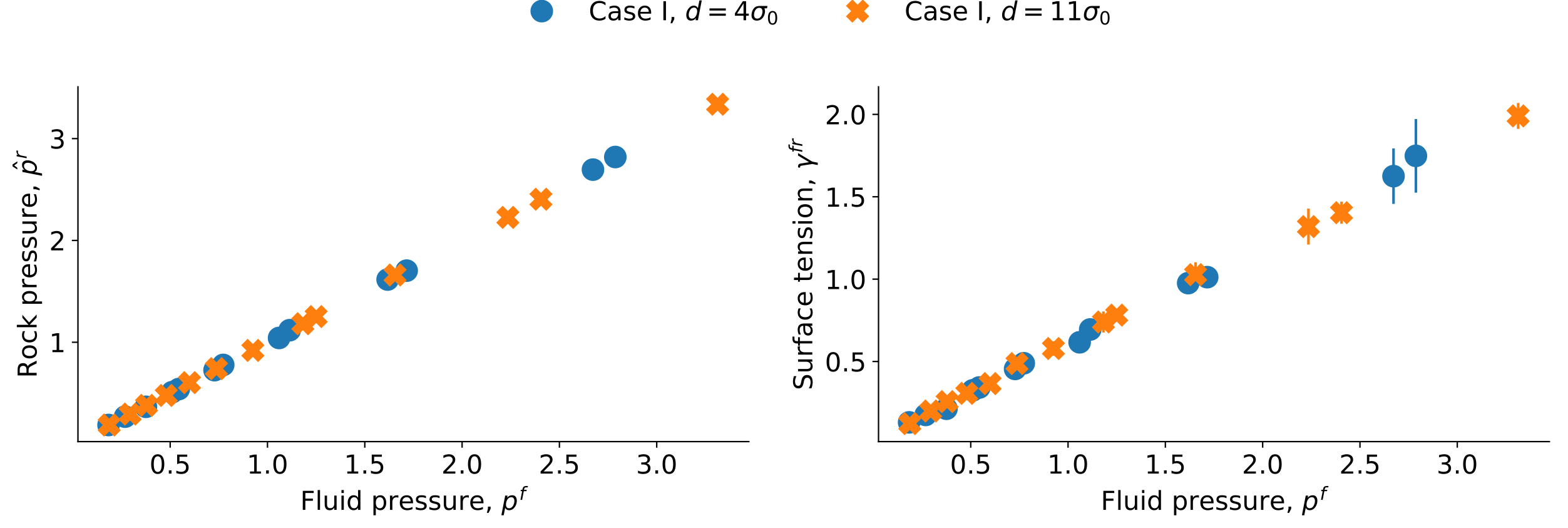}
  \caption{Fitted grain pressure $\hat{p}^r$ and surface tension $\gamma^{fr}$ as a function of pressure $p$ for a sphere (characteristic length $d=4\sigma_0$ and $d=11\sigma_0$).}
  \label{fig:caseI}
\end{figure}

The plots of $\hat{p}^{r}$ and $\gamma ^{fr}$ as functions of $p$ in region B are shown in figure  \ref{fig:caseI}. 
The values for $d=4\sigma_0$ and $d=11\sigma_0$ are given in the same plots.  We see that 
the plots fall on top of each other. This shows that the integral pressure and the surface tension are independent of the distance $d$ in the interval considered, and therefore above the lower limit $4\sigma_0 < d $. 
If confinement effects were essential, 
we would expect that $\hat{p}^{r}$ and $\gamma ^{fr}$ were functions of the distance $d$ between the surfaces of the spheres. When the value of $d$ decreases below $4\sigma_0$, deviations may arise, for instance due to contributions from the disjoining pressure. Such a contribution is expected to vary with the surface area, and increase as the distance between interfaces become shorter. In plots like figure  \ref{fig:caseI}, we may see this as a decrease in the surface tension. 

\subsection{Case II. Lattice of spherical grains. Equilibrium}

\begin{figure}
  \centering
  \includegraphics[width=0.85\linewidth]{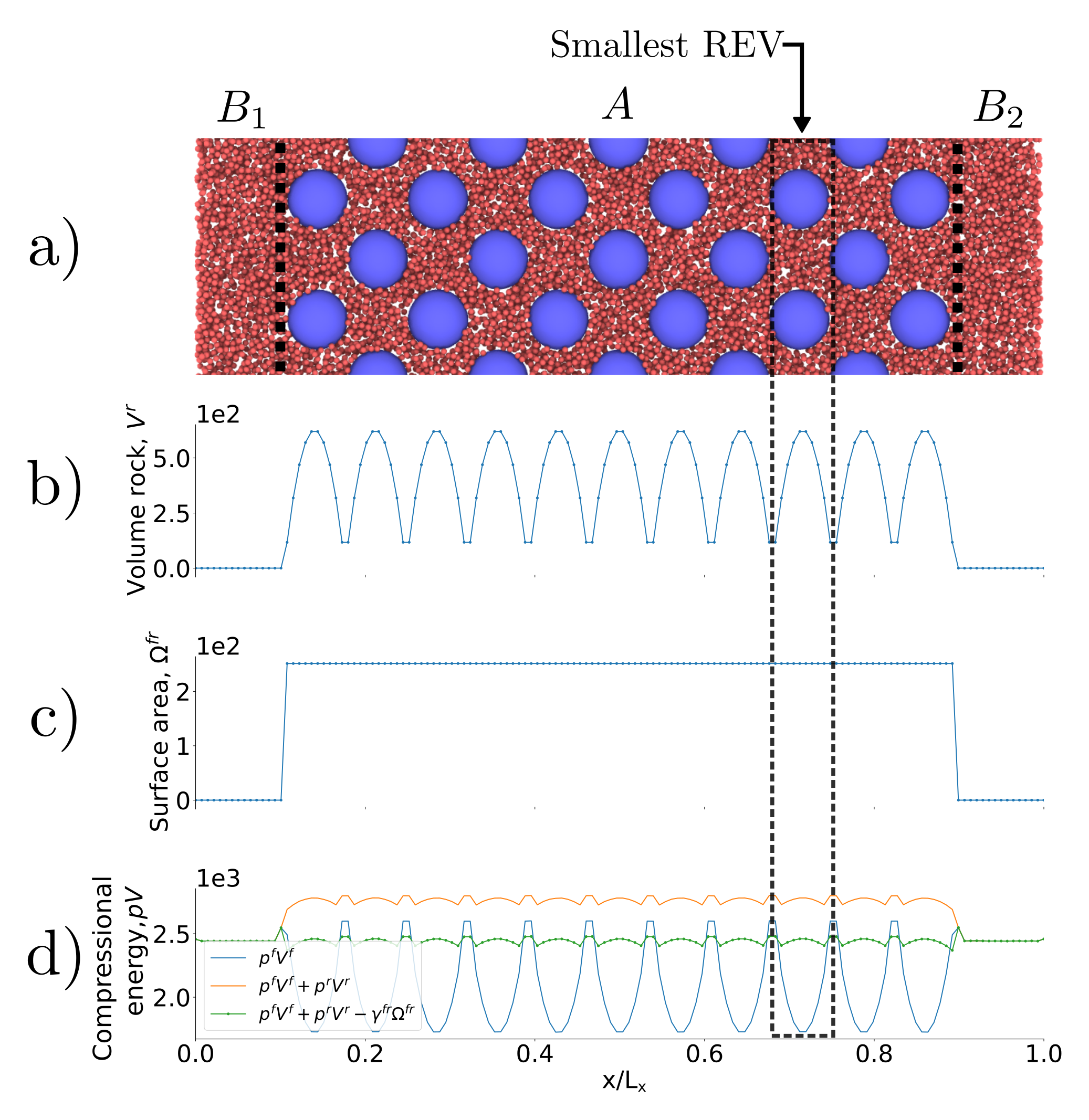}
  \caption{\textbf{a)} Illustration of case II, a lattice spherical grain surrounded by a fluid phase with $a=20\sigma_0$. \textbf{b)} Volume of grain, $V^r$, \textbf{c)} surface area $\Omega^{fr}$ and \textbf{d)} compressional energy $pV$ as a function of the $x$-axis of the simulation box. The smallest REV is half a unit cell.}
  \label{fig:several_spheres}
\end{figure}

Consider next the lattice of spherical grains, illustrated in figure \ref{fig:several_spheres}a. Figure \ref{fig:several_spheres}b and \ref{fig:several_spheres}c give the variation in the volume of the porous medium $V_l^r$ and surface area, $\Omega^{fr}$, along the $x$-axis. 

When the REV in region A is properly chosen, we know that $p^\rev=p$. In equilibrium, the pressure of the REV is constant in the bulk liquid phases, in regions $B_{1}$ or $B_{2}$, where $p$ is the pressure of the fluid in region B. In order to obtain $pV^{\rev}$ in region A, we sum $p_{l}V_{l}$ over all the layers that make up the REV, and obtain
\begin{equation}
  pV^\rev=\sum_{l\in \rev}p_{l}^{f}V_{l}^{f}+\hat{p}^{r}\sum_{l\in \rev}V_ l^r-\gamma
  ^{fr}\sum_{l\in \rev}\Omega _{l}^{fr},  \label{R.3}
\end{equation}
To proceed, we find first the values of all the elements in this equation, except $\hat{p}^{r}$\ and $\gamma ^{fr}$. The values of $\hat{p}^{r}$ and $\gamma ^{fr}$ are then chosen such that equation \ref{R.3} holds. Using these fitted values, we next calculated $\hat{p}_{l}V_{l}$ of each layer using 
\begin{equation}
  p_{l}V_{l}=p_{l}^{f}V_{l}^{f}+\hat{p}^{r}V_{l}^{r}-\gamma
  ^{fr}\Omega _{l}^{fr}  \label{R.4}
\end{equation}
The contributions to the compressional energy in this equation are shown in three stages in figure \ref{fig:several_spheres}d: (1) bulk fluid contribution $p_{l}^{f}V_{l}^{f}$, (2) bulk fluid and grain contribution $p_{l}^{f}V_{l}^{f}+\hat{p}^{r}V_{l}^{r}$ and (3) the total compressional energy, $p_{l}V_{l}=p_{l}^{f}V_{l}^{f}+\hat{p}^{r}V_{l}^{r}-\gamma^{fr}\Omega^{fr}$. Figure \ref{fig:several_spheres}d shows clearly that the bulk contribution is largest, as is expected. However, the surface energy is significant. 
From figure \ref{fig:several_spheres}b it follows that a proper choice of the REV is half a unit cell, because all REVs  are then identical, (except the REVs at the boundaries). The integral over $p_lV_{l}$ in these REVs is the same and equal to $pV^\rev$. The layers $l$ are smaller than the REV and as a consequence $\hat{p}_{l}V_{l}$ will fluctuate, a fluctuation that is seen in   figure \ref{fig:several_spheres}d. 

\begin{figure}
  \centering
  \includegraphics[width=0.85\linewidth]{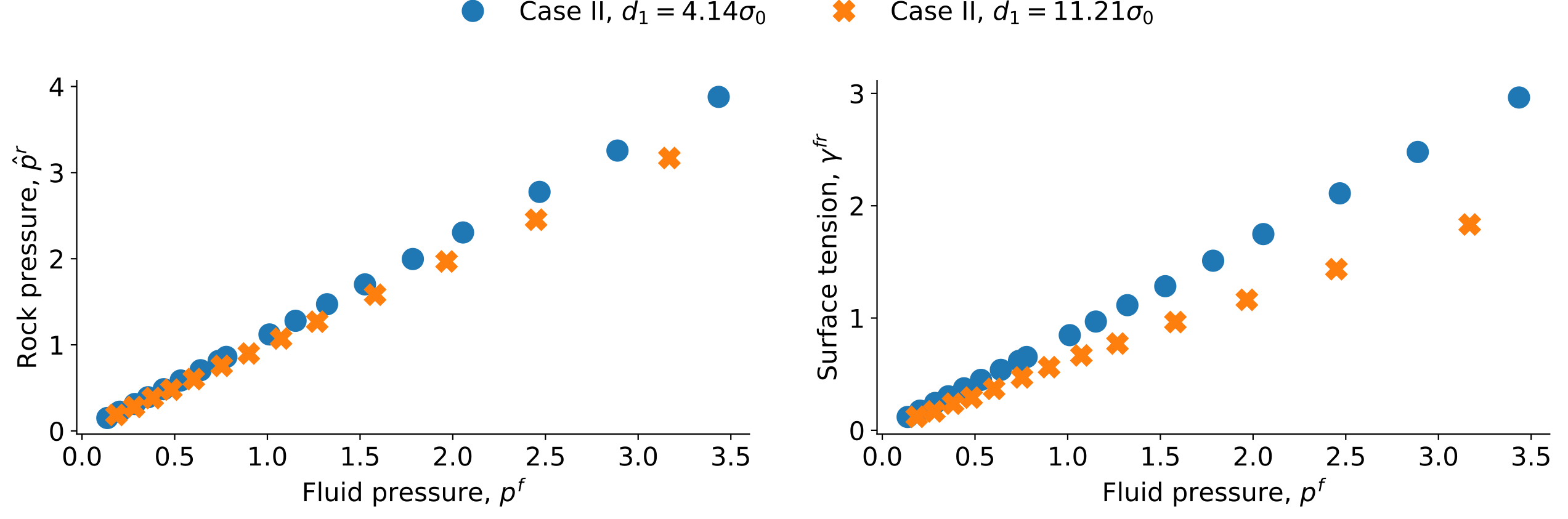}
  \caption{Fitted grain pressure $\hat{p}^r$ and surface tension $\gamma^{fr}$ as a function of pressure $p$ for the lattice of spheres (characteristic length $d_1=4.14\sigma_0$ and $d_1=11.21\sigma_0$). }
  \label{fig:caseII}
\end{figure}

The values for $\hat{p}^{r}$ and $\gamma ^{fr}$ are shown as function of $p$ for Case II in figure \ref{fig:caseII} for $d_1=4.14\sigma_0$ and $d_1=11.21 \sigma_0$. We see now a systematic difference between the values of  $\hat{p}^{r}$ and $\gamma ^{fr}$  in the two cases. The intergral pressure and the surface tension increases as the distance between the grains decreases.
The difference in one set can be estimated from the other. Say, for a difference in surface tension $\Delta \gamma^{fr}$  we obtain for the same fluid pressure from equation \ref{eq:newrelation}, a difference in integral pressure of  $\Delta \hat{p}^r = 3 \Delta \gamma^{fr}/R$. This is nearly what we find in by comparing the lines in figure \ref{fig:several_spheres}, the lines can be predicted from one another using $R=6.5\sigma_0$  while the value in figure \ref{fig:rdf} is $R=5.5\sigma_0$. The difference must be due to the disjoining pressure. Its distribution is not spherically symmetric, which may explain the difference between $6.5\sigma_0$ and $5.5\sigma_0$. 

\begin{figure}
  \centering
  \includegraphics[width=0.85\linewidth]{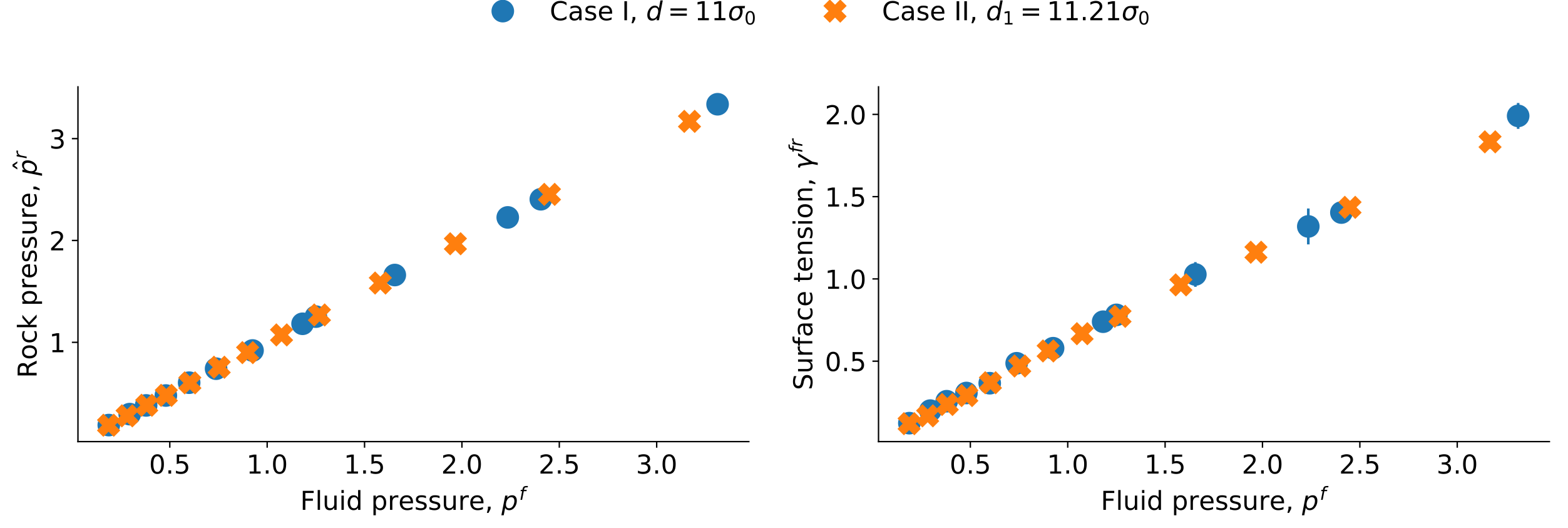}
  \caption{Fitted grain pressure $\hat{p}^r$ and surface tension $\gamma^{fr}$ as a function of pressure $p$ for the sphere (characteristic length $d=11\sigma_0$) and a lattice of spheres (characteristic length $d_1=11.21\sigma_0$).}
  \label{fig:compare}
\end{figure}

The results should be the same as for Case I for the larger distance, and indeed that is found, cf. figure \ref{fig:compare}. 
As the distance between the grain surfaces increases, we expect the dependence of confinement to disappear, and this documented by figure \ref{fig:compare} where the two Cases are shown with distances $d=11\sigma_0$ and $d_1=11.21\sigma_0$, respectively. The curves for the single grain and lattice of grains overlap. 

Our analysis therefore shows that the pressure inside grains in a fcc lattice and the surface tension, depends in particular on the distances between the surfaces of the spheres, including on their periodic replicas. 
A procedure has been developed to find the pressure of a REV, from information of the (equilibrium) values of  $\hat{p}^{r}$ and $\gamma ^{fr}$ as a function of $p$. It has been documented for particular nano-porous medium, but is likely to hold for other lattices.

\subsection{The representative elementary volume.}

The knowledge gained above on the various pressures at equilibrium, can be used to first construct the REV and next the pressure variation away from equilibrium. Following Kjelstrup et al. \cite{Kjelstrup2018a}, we assume then that there is \textit{local} equilibrium in the REV. The size of the REV was established above, as the minimum size that cover the complete range of potential interactions available in the system. To find a REV-property, we need to sample the whole space of possible interactions.  

The thickness of the REV is therefore larger than the layer thickness used in the simulations. To show how a REV-property is determined from the layer-property, we consider again the compressional energies of each layer, but now for a system in a pressure gradient.  

\begin{figure}
  \centering
  \includegraphics[width=0.7\linewidth]{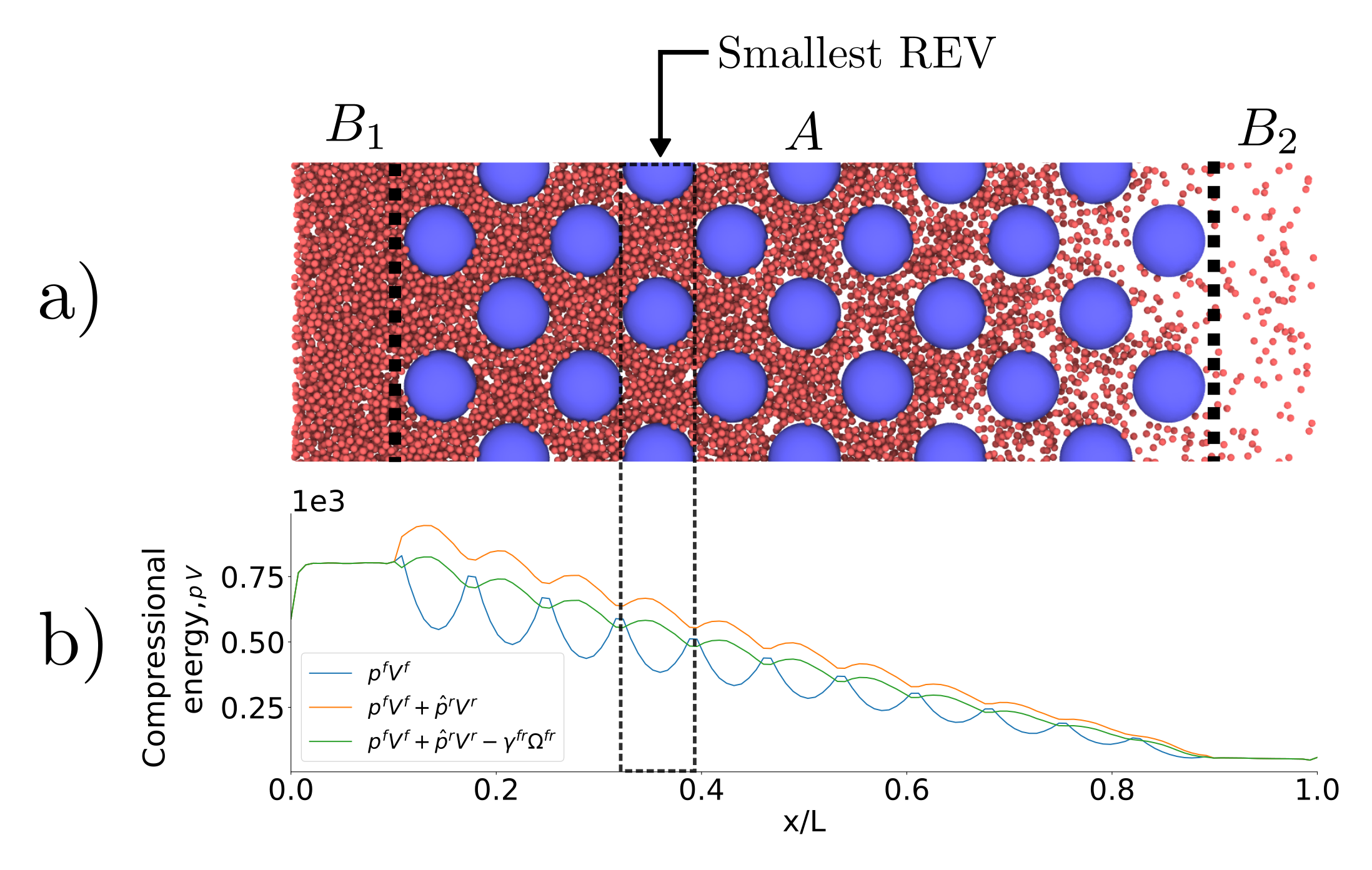}
  \caption{\textbf{a)} Illustration of case II in a pressure gradient. \textbf{b)} Compressional energy $pV$ variation across the system.}
  \label{fig:pressure_gradient}
\end{figure}

\begin{figure}
  \centering
  \includegraphics[width=0.7\linewidth]{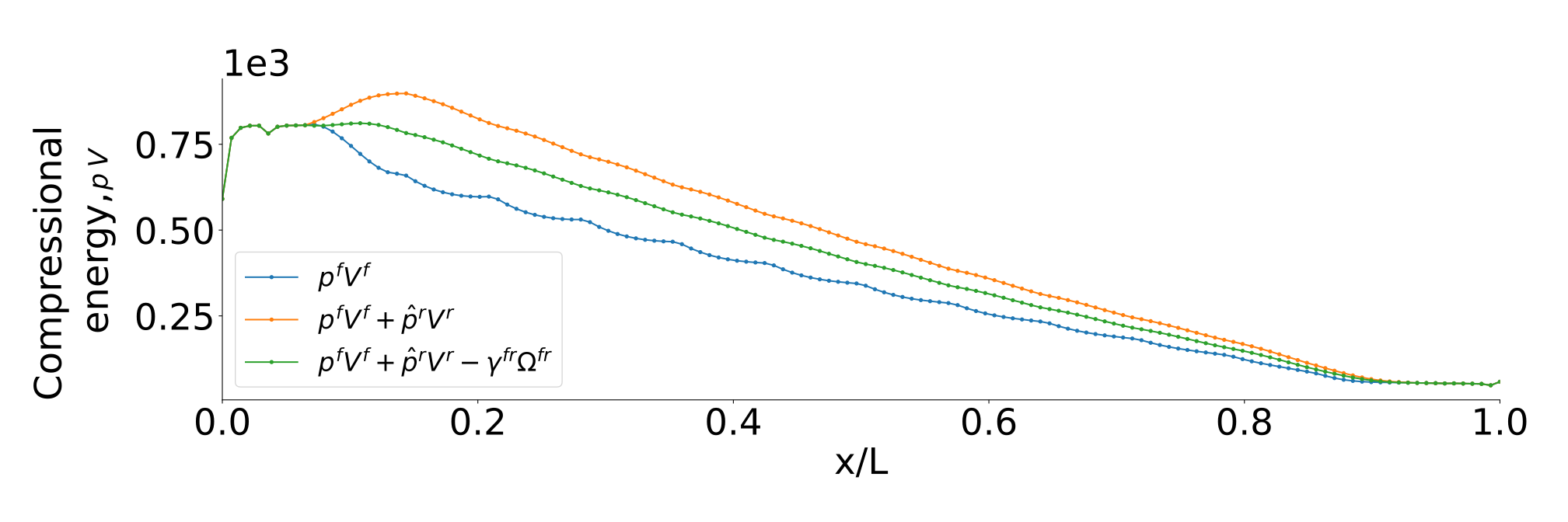}
  \caption{Compressional energy $pV$ variation across the system smoothed over the representary elementary volume.}
  \label{fig:pressure_gradient_smooth}
\end{figure}

In the analysis we used the fcc lattice with lattice parameter $a=20\sigma_0$. The volume of the grain, $V^r$, and the surface area, $\Omega^{fr}$, varied of course in the exact same way as in figure \ref{fig:several_spheres}b and c. 
The pressure gradient was generated as explained in section \ref{sec:neq}. The pressure difference between the external reservoirs B$_1$ and B$_2$ was large, giving a gradient with order of magnitude $10^{12}$ bar/m. The fluid on the left hand side is liquid-like, while the fluid on the right-hand side is gas-like. 

Figure \ref{fig:pressure_gradient} illustrates the system in the pressure gradient, where subfigure \ref{fig:pressure_gradient}b shows the compressional energy, $pV$, along the $x$-axis. The dip in the pressure close to $x=0$ is caused by the reflecting particle boundary, cf. section \ref{sec:neq}. The reflecting particle boundary introduces a surface between the high fluid on the left side and the low pressure on the right side.

The smallest REV as obtained in the analysis at equilibrium is indicated in the figure. 
Within a REV, the variables fluctuate, as clearly shown in the equilibrium results, figure \ref{fig:several_spheres}. 

In order to compute a REV variable, we follow the procedure described by Kjelstrup et al \cite{Kjelstrup2018a} and choose a layer as a reference point. We then compute the moving average using five layers, two to the left, two to the right and the central layer.
Moving one layer down the gradient,  we repeat the procedure, and in this manner we obtain the property variation on the REV scale. The results of the simulation gave, for each individual layer, $p_l^{l}V_{l}^{l}$, as plotted in figure \ref{fig:pressure_gradient}b. The profile created by the REV-centers is shown in figure \ref{fig:pressure_gradient_smooth}. We see a smooth linear profile (central curve) as one would expect from the boundary conditions that are imposed on the system. Some traces of oscillation are still left in the separate contributions to the total compressional energy. The total energy is constant in regions A, as expected.

\subsection{Case II. Lattice of spherical grains. Non-equilibrium}

The results shown in figure \ref{fig:pressure_gradient_smooth} mean that the integral as well as the differential pressures vary through the system.  The relation between $p$ and $\hat{p}^{r}$, $\gamma ^{fr}$ in the REV is the same as in figure \ref{fig:compare} at local equilibrium. We can therefore calculate $\hat{p}_{l}^{r}$ and $\gamma_{l}^{fr}$ for each layer from this figure for each REV.   

The compressional energy $p_{l}V_{l}$, as determined from
\begin{equation}
  p_{l}V_{l}=p_{l}^{f}V_{l}^{f}+\hat{p}%
  _{l}^{r}V_{l}^{r}-\gamma _{l}^{fr}\Omega _{l}^{fr}  \label{NE}
\end{equation}
gives the total compressional energy $p_{l}V_{l}$, and finally the REV compressional energy. The last property shows an essentially linear variation from the value in one bulk fluid to the value in the other bulk fluid, cf. figure \ref{fig:pressure_gradient_smooth}. The fluctuations around the mean value were discussed above. The REV should be large enough to eliminate these. 

We have found that a nano-porous medium has an increased number of variables. We need consider the pressures in the fluid and the solid phases, as well as the surface tension between the fluid and the solid. When one reduces the size of a thermodynamic system to the nano-meter size, the pressures and the surface tensions become dependent on the size of the system. An important observation is then that there are two relevant pressures rather than one. Hill \cite{Hill1964} called them the integral and the differential pressure, respectively. It is maybe surprising that the simple virial expression works so well for all pressure calculations in a fluid, but we have found that it can be used. This means that we will next be able to study transport processes, where the external pressure difference is a driving force. The method, to compute the mechanical force intrinsic to the porous medium may open interesting new possibilities to study the effects that are characteristic for porous media.   

In a macro-scale description the so-called representative elementary volume (REV) is essential. This makes it possible to obtain thermodynamic variables on this scale. We have here discussed how the fact, that the macro-scale pressure is constant in equilibrium, makes it possible to obtain the integral pressure in the solid, as well as the surface tension of the REV. An observation which confirms the soundness of the procedure is that we recover that the resulting differential pressure in the solid satisfies Young-Laplace's law. The existence of a REV for systems on the nano-scale supplements the REV that can be defined for pores of micrometer dimension \cite{Kjelstrup2018a}. There is no conflict between the levels of description as they merge in the thermodynamic limit. The REV, as defined in the present work, may allow us to develop a non-equilibrium thermodynamic theory for the nano-scale.  

\section{Conclusions}

The following conclusions can be drawn from the above studies 
\begin{itemize}
    \item We have obtained the first support for a new way to compute the pressure in a nano-porous medium. 
     The integral pressure is defined by the grand potential. The definition applies to the thermodynamic limit, as well as to systems which are small according to the definition of Hill \cite{Hill1964}.  
    \item It follows that nano-porous media need two pressures in their description, the integral and the differential pressure. This is new knowledge in the context of nano-porous media
    \item For a spherical rock particle of radius $R$, we derive a relation between the integral and the differential pressure in terms of the surface tension, $\hat{p}^r-p^r = {\gamma}/{R}$.  Their difference is non-negligible in the cases where Young-Laplace's law applies.
    \item To illustrate the calculation, we have constructed a system with a single fluid.  The rock pressure and the surface tension are constant throughout the porous medium at equilibrium. The assumptions were confirmed for a porosity change from $\phi=0.74$ to $0.92$, for a REV with minimum size of half a unit cell. 
    \item From the assumption of local equilibrium, we can find the pressure internal to a REV of the porous medium, under non-equilibrium conditions, and a continuous variation in the pressure on a macro-scale. The procedure to use a virtual (sister) path to find the variable in question, may apply also to other variables than the pressure  \cite{Kjelstrup2018a}. 
\end{itemize}
To obtain these conclusions, we have used molecular dynamics simulations of a single spherical grain in a pore and then for face-centered lattice of spherical grains in a pore. This tool is irreplaceable in its ability to test assumptions made in the theory. The simulations were used here to compute the integral rock pressure and the surface tension, as well as the pressure of the representative volume, and through this to develop a procedure for porous media pressure calculations. 

Only one fluid has been studied here. The situation is expected to be much more complicated with two-phase flow and an amorphous medium. Nevertheless, we believe that this first step has given useful information  for the work to follow. We shall continue to use the grand potential for the more complicated cases, in work towards a non-equilibrium thermodynamic theory for the nano-scale.   

\section{Acknowledgement}
The calculation power was granted by The Norwegian Metacenter of Computational Science (NOTUR). Thanks to the Research Council of Norway through its Centres of Excellence funding scheme, project number 262644, PoreLab.

\section*{Appendix} 

This appendix describes an alternative way to derive equation \ref{eq:equil} in section \ref{case1} and an integral form of the Young-Laplace law. Inserting equation \ref{eq:p_hatV} into equation \ref{eq:p_hill} gives
\begin{equation}
  p(V) = \frac{\partial \left( \hat{p}^f V^f\right) }{\partial V}
+  \frac{\partial \left( \hat{p}^r V^r\right) }{\partial V}
- \frac{\partial \left( \hat{\gamma}^{fr} \Omega ^{fr}\right) }{\partial V}
  \label{eq:A1}
\end{equation}
From Hill \cite{Hill1964} we can express the differential pressures and the interfacial tension as
\begin{subequations}
\begin{align}
p^f &=\frac{\partial \left( \hat{p}^f V^f\right) }{\partial V^f} \\
p^r &=\frac{\partial \left( \hat{p}^r V^f\right) }{\partial V^r} \\
\gamma^{fr} &=\frac{\partial \left( \hat{\gamma}^{fr} \Omega^{fr}\right) }{\partial \Omega^{fr}}
\end{align}
\label{eq:A5}
\end{subequations}
The porous medium is characterized by a porosity,
\begin{equation}
\phi=\frac{V^f}{V}
\end{equation}
If the change in volume implied by the differentiation in equation \ref{eq:A1} is performed at constant porosity (proportional swelling of the two phases)\footnote{This may be the only way to do the differentiation consistent with Euler homogeneity}, we get
\begin{equation}
\frac{\partial}{\partial V} = \phi \frac{\partial}{\partial V^f}
 = (1-\phi) \frac{\partial}{\partial V^r} = (1-\phi) \frac{2}{R} \frac{\partial}{\partial \Omega^{fr}}
\end{equation}
and
\begin{equation}
  p = \phi p^f + (1-\phi)\left(p^r - \frac{2 \gamma^{fr}}{R}\right)
 \label{eq:p_hill_contracted}
\end{equation}
In the limit $\phi \rightarrow 1$ $p \rightarrow p^f$ as one should expect. However, Equation \ref{eq:p_hill_contracted} involves only differential pressures. We can then apply Young-Laplace's law, $p^r=p^f+2\gamma^{fr} / R$, to the last term and find that
\begin{equation}
    p=\phi p^f + (1-\phi) p^f = p^f,
\end{equation}
which is equation \ref{eq:equil}.

Using the concept of porosity, we can rearrange equation \ref{eq:p_hatV}:
\begin{equation}
\hat{p} = \phi \hat{p}^f + (1-\phi) \hat{p}^r - (1-\phi)\frac{3 \hat{\gamma}^{fr}}{R}
\end{equation}
This suggests an alternative to Young-Laplace's law on the integral level:
\begin{equation}
\hat{p}^r = \hat{p}^f + \frac{3 \hat{\gamma}^{fr}}{R}
\end{equation}

\bibliographystyle{ieeetr}
\bibliography{paper1.bib}

\end{document}